\documentclass{article}

\usepackage{arxiv}

\usepackage[utf8]{inputenc} 
\usepackage[T1]{fontenc}    
\usepackage{hyperref}       
\usepackage{url}            
\usepackage{booktabs}       
\usepackage{amsfonts}       
\usepackage{nicefrac}       
\usepackage{microtype}      
\usepackage{lipsum}
\usepackage{subfigure}
\usepackage{graphicx}
\usepackage{natbib}
\RequirePackage{filecontents}

\title{Characterising Third Party Cookie Usage in the EU after GDPR}

\author{
  Xuehui Hu\\
  King's Collage London\\
  London, UK \\
  \texttt{xuehui.hu@kcl.ac.uk} \\
   \And
 Nishanth Sastry\\
  King's Collage London\\
  London, UK \\
  \texttt{nishanth.sastry@kcl.ac.uk} \\
}
\pdfoutput=1
\begin{document}
\maketitle

\begin{abstract}
{
The recently introduced General Data Protection Regulation (GDPR) requires that when obtaining information online that could be used to identify individuals, their consents must be obtained. Among other things, this affects many common forms of cookies, and users in the EU have been presented with notices asking their approvals for data collection. This paper examines the prevalence of third party cookies before and after GDPR by using two datasets: accesses to top 500 websites according to \texttt{Alexa.com}, and weekly data of cookies placed in users' browsers by websites accessed by 16 UK and China users across one year. 

We find that on average the number of third parties dropped by more than 10\% after GDPR, but when we examine real users' browsing histories over a year, we find that there is no material reduction in long-term numbers of third party cookies, suggesting that users are not making use of the choices offered by GDPR for increased privacy. Also, among websites which offer users a choice in whether and how they are tracked, accepting the default choices typically ends up storing more cookies on average than on websites which provide a notice of cookies stored but without giving users a choice of which cookies, or those that do not provide a cookie notice at all. We also find that top non-EU websites have fewer cookie notices, suggesting higher levels of tracking when visiting international sites. Our findings have deep implications both for understanding compliance with GDPR as well as understanding the evolution of tracking on the web.
}
\end{abstract}


\section{Introduction}

\begin{figure*}[!t]
\vspace{-2mm}
\centering
\subfigure[Levelled cookies setting in Forbes.com] {\includegraphics[width=2.2in]{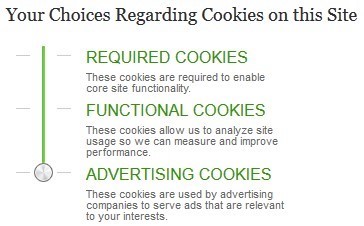}}
\subfigure[Detailed Cookie Table provided by LinkedIn.com] {\includegraphics[width=3.8in]{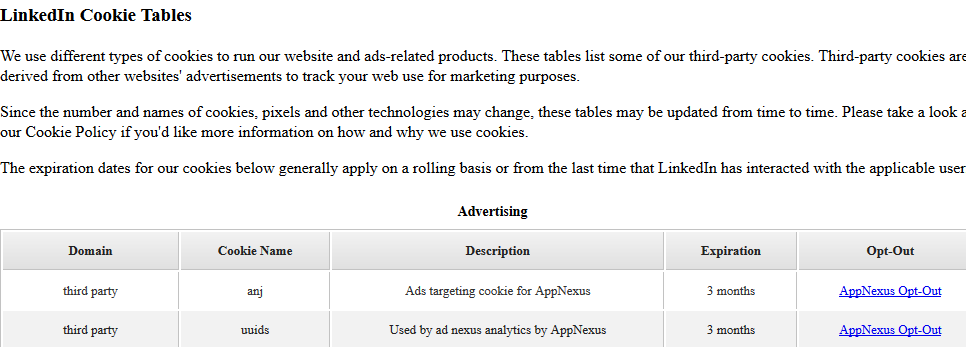}}
\subfigure[Office.com provides a cookie notice but no choice] {\includegraphics[width=5.5in]{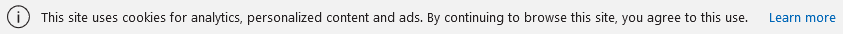}}
\vspace{-2mm}
\caption{Examples of cookie notices provided by website owners to EU users after GDPR came into effect (May 25, 2018)}
\label{cookie_linkedin}
\vspace{-4mm}
\end{figure*}
The General Data Protection Regulation (GDPR) is a sweeping regulation that came into effect on May 25, 2018 in the European Union (EU), to protect the online privacy of its residents~\cite{gdpr_site}. GDPR affects many aspects of personal data collection~\cite{gdpr_law_review}, although some argue that it does not go nearly far enough~\cite{gdpr_bigdata}. A central tenet of GDPR is that whenever personal data is collected about a user, it has to be done with the consent of the user. 

This notion of user consents has affected a large number of sites that have used various mechanisms including analytics, tracking, and targeted advertising to track users. Such websites are now required to inform users. Consent for cookies which can be used to identify a user uniquely is explicitly mentioned in Recital 30~\cite{GDPR_recital30}. 

The need to inform users has led to a large number of cookie notices to users. Different websites have adopted different practices as shown in Fig.~\ref{cookie_linkedin}. Some, such as Forbes and LinkedIn (Fig.~\ref{cookie_linkedin}~(a)~\&~(b)) have provided users with several choices, allowing them to select or unselect different options. Others, such as Office.com (Fig.~\ref{cookie_linkedin}~(c)) simply inform (without giving the user any choice) that user-specific cookies are being used, and this notice needs to be accepted if the website is accessed. The last option is not to issue any notice at all, because either no user-specific cookie has been used, or non-compliance of GDPR. Many websites appear to have chosen one of the first two options (cookie notice with or without choice) because GDPR non-compliance can attract fines of up to the higher of 20 Million Euros or 4\% of the turnover of a company\footnote{Art. 83(4) and 84(5) of the GDPR.  https://gdpr-info.eu/art-83-gdpr/}.

In this paper, we investigate GDPR cookie notices on two sets of websites. The first is the set of top sites according to Alexa Web Traffic Analysis. The second set comprises websites visited by real users in an ongoing study\footnote{All collected data have been obtained with agreement from participants and under Research Ethics Minimal Risk Registration process at our university 
to ensure the permissions of approvals relevant to this research (Ethics approval no.\ MRS-17\/18-6539)}. In both cases, we focus on so-called \emph{third party cookies}, i.e., cookies set not by the ``first party'' sites visited by the users,  but by other third parties used by the first party sites. For example, if a user visits a site that uses Google Analytics, a Google (Analytics) cookie is placed in the user's browser. Third party sites hold enormous power since they obtain a panoramic view of  a user's browsing history across different sites using the same third party.

We access these sets of websites from a vantage point in the EU, and obtain the following results:

\begin{enumerate}
\item Generally, websites which offer users a choice store \emph{more} third-party cookies (when users accept default options offered), than sites which do not give users a choice. Some websites appear to continue placing cookies that are used to track users even after they explicitly decline consent\footnote{Example screencast videos for such websites in Top500: https://bit.ly/2GnWrim}.
\item The number of third party cookies, as well as the manner of GDPR consent notices, vary across different categories of websites. Adult websites are the least likely to offer GDPR consent and choices, but also appear to contain fewer third party cookies, likely because several common third parties such as Facebook and DoubleClick do not work with adult sites. In contrast, news websites have the highest number of third parties, and also provide more cookie consent notices.
\item The prevalence of third-party cookies appears to differ across countries: Nearly 90\% (66\%) websites in the \texttt{Alexa.com} Top 100 in China (USA) do not issue any third party cookie notices, or provide no choice to  users on the manner of tracking. 
\item On average, the number of third-party cookies from UK  websites drops by 10\% after May 25, 2018, suggesting that GDPR has been successful and sites are complying with the regulation. However, this reduction appears to not be reflected in real users' browsing histories, and third party cookie numbers in 2019 show little change since before GDPR. 
\end{enumerate}
\vspace{-4mm}



\section{Datasets}

Our results are based on two datasets. The first dataset focuses on the top websites, i.e., those which obtain the maximum amount of traffic according to \texttt{Alexa.com}~\cite{Alexa_site}. We first analysed the top 100 sites in the UK one week before and one week after the introduction of GDPR (May 25 2018), focusing on differences in cookie numbers. In addition, we manually examine the types of cookie notices served by the top 500 websites in the UK after GDPR has been introduced.

The second dataset is obtained from a study in which anonymised browser histories are being collected weekly from 15 users (9 in the UK; 6 in China). We have instrumented the browsers of these users using a modified version of a browser plugin, Lightbeam~\cite{firefox_2012} which runs also on Google Chrome. Our plugins collect information about both the first party websites they visit, as well as the third party cookies placed as a result of visiting those first party sites. Altogether these users have visited around 15k first-party websites across the year, which have led to over 187k third-party domains from which cookies are  placed on their computers (Table \ref{table:table1}). We focus on the UK users who have visited around 8416 websites and have cookies from nearly 113K third-party domains. 
\begin{table}[ht]
  \begin{center}
    \begin{tabular}{l|c|c} 
   \normalsize   \textbf{User Group} & \normalsize \textbf{No.\ $1^{st}$ party sites} & \normalsize \textbf{No.\ $3^{rd}$ party cookies}\\
      \hline
     UK Users & 8416 & 113,003\\
      CN Users & 6144 &74,313\\
     Total & 14827 & 187,316\\\hline
      
    \end{tabular}
  \end{center}
  \caption{Data collected from Jan. 2018 to Jan. 2019}
       \vspace{-5mm}
    \label{table:table1}
  \vspace{-6mm}
\end{table}

\section{GDPR notices in Alexa top websites}
We first study  GDPR cookie notices in popular websites. 
Our study comprises three steps. First we capture cookies one week before and one week after GDPR comes into effect, among the \texttt{Alexa.com} Top 100 sites in the UK, which, as a current member of the EU, is subject to GDPR. Next we compare  UK cookie notices after GDPR was introduced, with those from outside the EU, taking USA and China as examplar non-EU countries, and also using \texttt{Alexa.com}'s global lists of top sites in various important categories of the web, such as shopping and technology. We then manually examine the different kinds of cookie notices among the top 500 websites in the UK, and discuss the impact  on tracking and GDPR compliance.



\subsection{Cookie notices among Alexa Top 100 sites} \label{default_setting}
\begin{figure*}[ht]
\centering
\subfigure[Cookie notice with choice (42 sites)] {\includegraphics[width=2.2in]{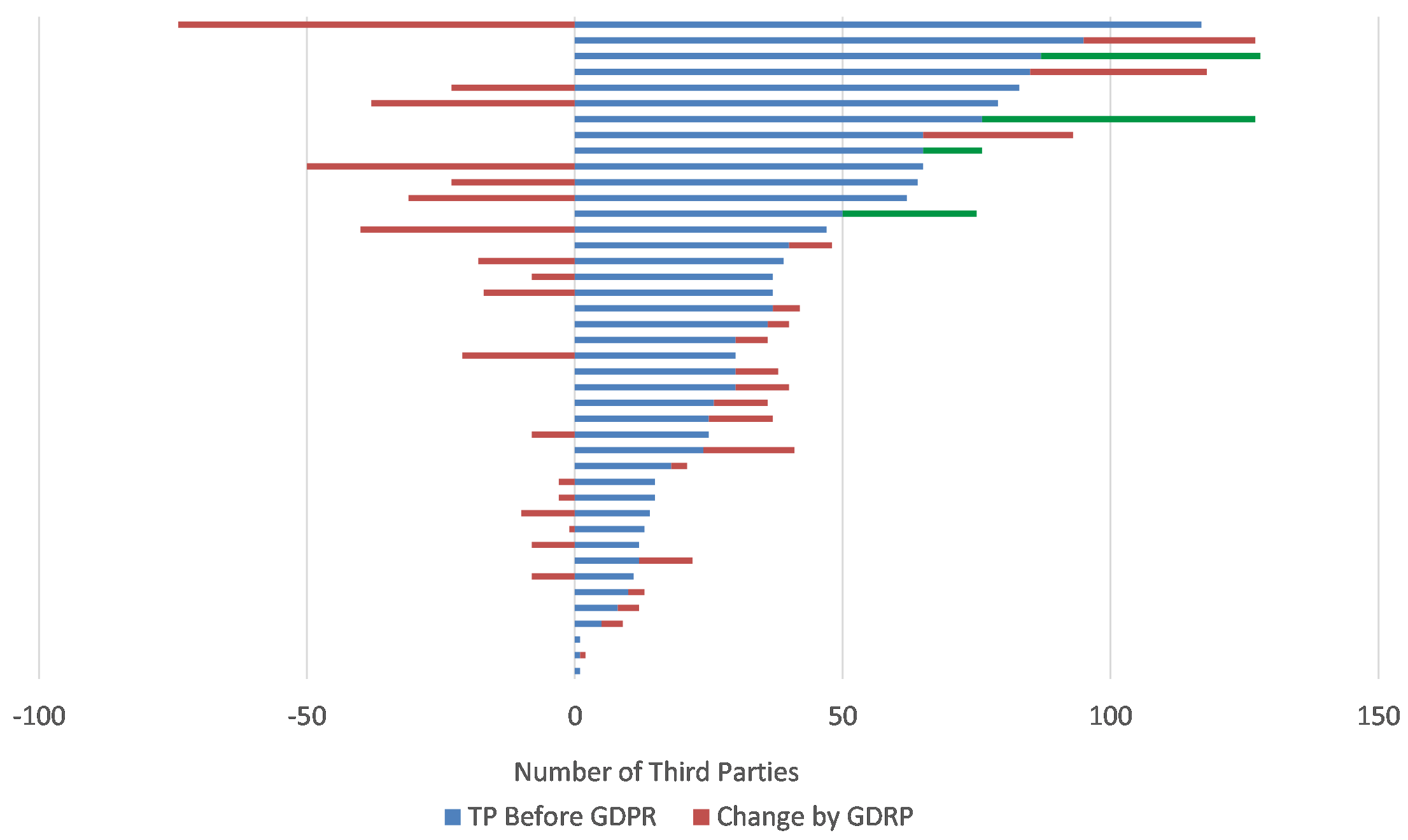}}
\subfigure[Cookie notice no choice (35 sites)] {\includegraphics[width=1.9in]{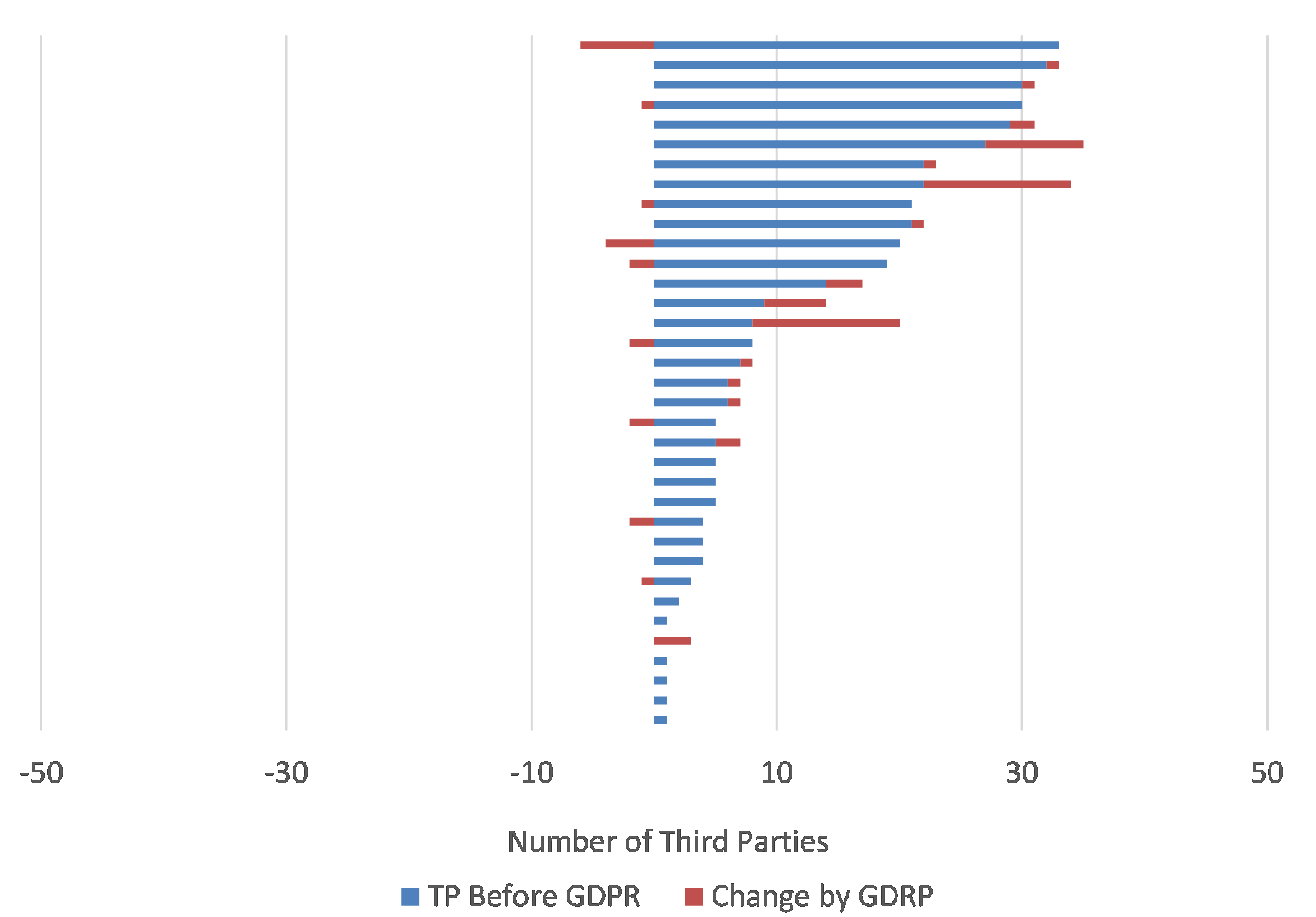}}
\subfigure[No cookie notice (23 sites)] {\includegraphics[width=2in]{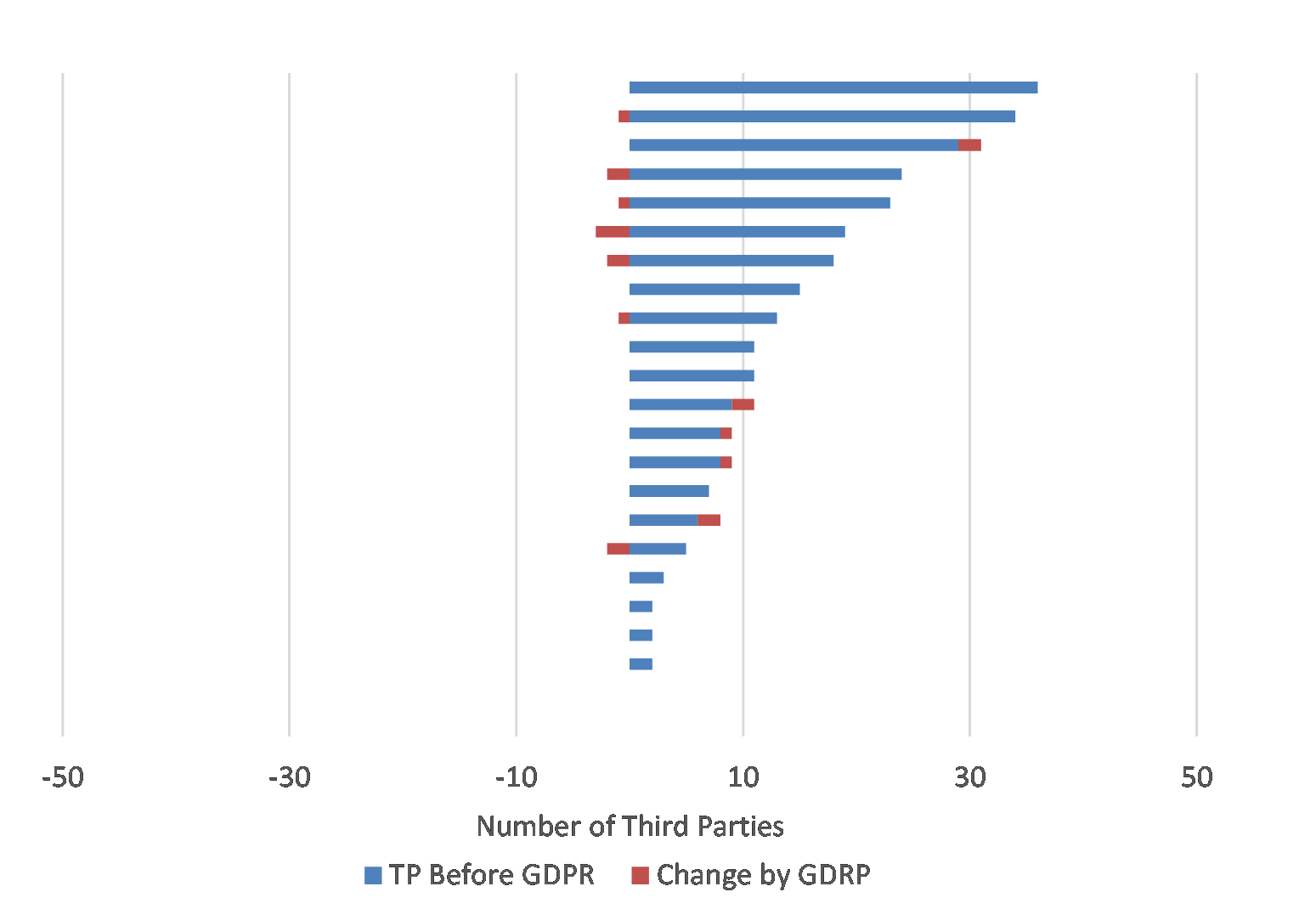}
}
\caption{The changes on the number of third-party cookies of Alexa Top100 Websites (one week before and after GDPR), if the default choice is accepted. Each horizontal line denotes a site, totally 100 lines across three subgraphs. For each site, blue shows the number of third-party cookies served before GDPR, and red the \emph{change} in the number of cookies after GDPR. Three categories are observed: (a) Sites which serve users with cookie notices. (Green indicate sites which store cookies even if users explicitly opt out) (b) Sites which serve cookie notices but offer no choice to users. (c) Sites which serve no notices after GDPR.}
\label{fig:tp}
\end{figure*}
 \begin{figure}[!ht]
\centering
\includegraphics[width=0.62\textwidth]{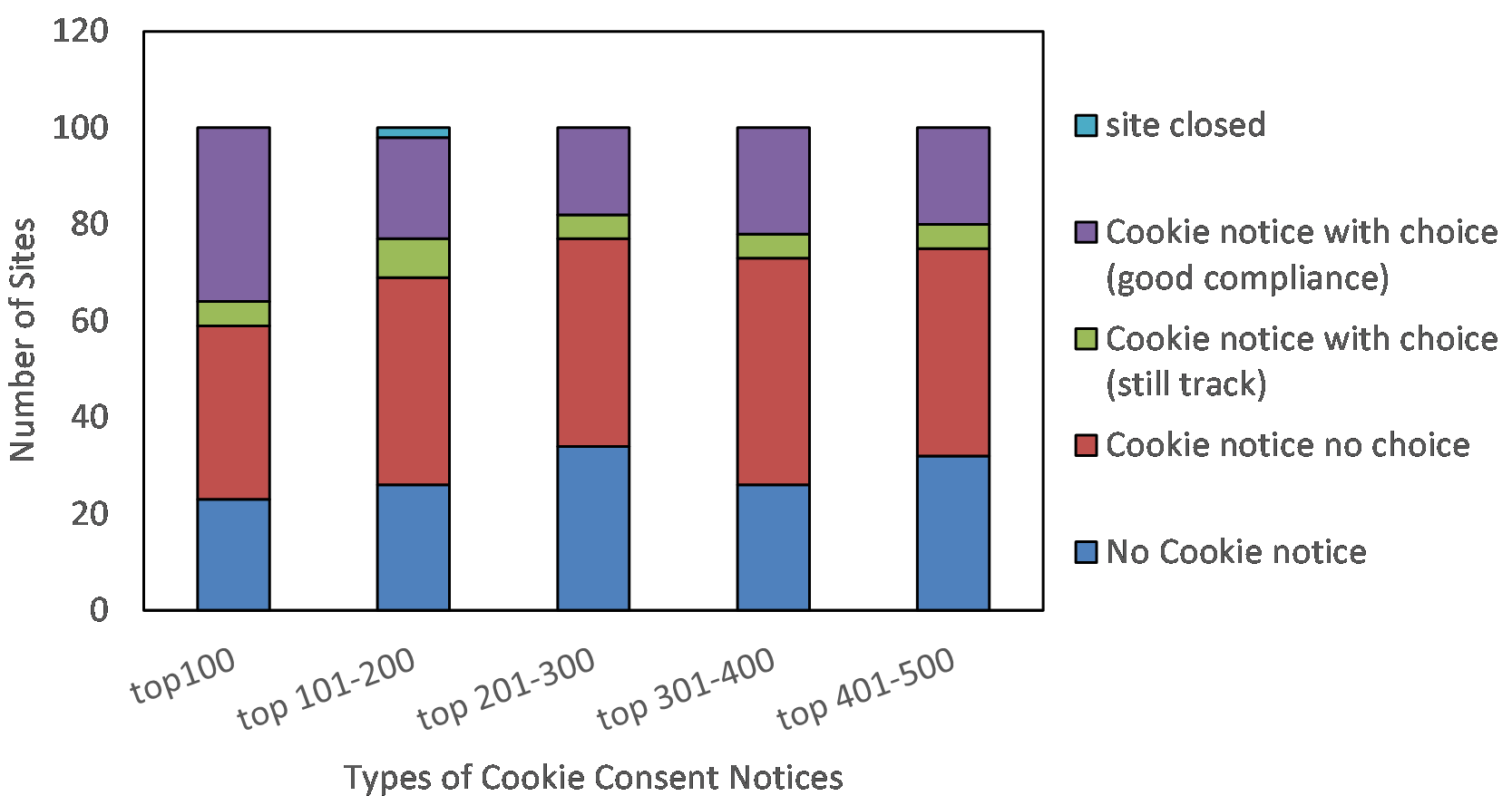}
\caption{\label{fig:top500_cookietype} Detailed study of UK top500 sites' Cookie Types.}
\vspace{-6mm}
\end{figure}

  After $25^{th}$ May in 2018, websites started to pop up cookie notices to users before data from them is collected.  Generally, there are three types of cookie notices: The first one is that the website owner provides users with a  privacy choice of opting out from the data sharing, e.g., Forbes and LinkedIn (Fig.~\ref{cookie_linkedin}~(a)~\&~(b)). Other examples include Reddit, Twitter and Amazon. 
 The second kind of websites includes vendors  that provide a notice of cookie collection but they do not offer a way to change the setting, e.g., Office.com (Fig.~\ref{cookie_linkedin}~(c)). Essentially, the user has to choose between using the website with cookies being used, and not using the website at all. The final kind of websites provide no cookie collection notice. A handful of websites also stop their business and support for European users. This includes several prominent non-EU sites such as LAtimes.com, ChicagoTribune.com, QQ.com, Unroll.me, etc.
 
 Fig.~\ref{fig:tp} studies GDPR cookie notices of the \texttt{Alexa.com} Top 100 websites in the UK. Nearly 80\% of these sites display some form of cookie notice (Fig.~\ref{fig:tp}~(a)~\&~(b)), and half of all collected websites provide an option on whether to receive personalised ads or not (Fig.~\ref{fig:tp}~(a)). When the websites provide a choice, we accept the default settings and observe the number of cookies stored\footnote{Note that some of the cookies stored are simply to note the fact that the cookie notice has been served and accepted. We discard these cookies from our counts.}. 22 websites in the top 100 do not serve any cookie notice.
 
  As expected, GDPR appears to have had an effect on the \emph{number} of third party cookies immediately after the law came into effect. Amongst websites which allow users to set their choices (Fig.~\ref{fig:tp}~(a)), the average number of third party cookies dropped from 34 to 28; websites which show a cookie notice but provide no choice in the matter (Fig.~\ref{fig:tp}~(b)) show a minor reduction from 16 cookies on average before GDPR to 15 after; those which do not issue cookie notices (Fig.~\ref{fig:tp}~(c)) show no change, with an average of 13 third party cookies before and after GDPR. 
  
 \noindent\textbf{Degree of GDPR compliance}: 
   It is interesting and notable that websites which appear to be transparent and offer users a choice (Fig.~\ref{fig:tp}~(a)) store \emph{more} cookies (avg. 28) when the default option is accepted, than those which provide no choice (avg. 15).  Similarly, several websites which offer an option seem to have used the opportunity to \emph{increase} the number of third-party cookies  (Red lines on the positive side of Fig.~\ref{fig:tp}~(a)).
 Examining manually, we see that websites which do not serve cookie notices \emph{use some of the same third party trackers (e.g., Google Analytics or Facebook cookies) which are found among websites that do serve notices}, which  suggests that perhaps such websites \emph{should} be serving cookie notices and asking for user consent, or  could be not compliant with GDPR. 
 
 Furthermore, in our manual examination of websites that do provide users with a choice, we see cases where tracking cookies are being placed even after opting out of tracking and personalisation (i.e., even when we choose non-default choices that maximise privacy), highly indicative of GDPR non-compliance (See footnote 3, Pg.~1).    Fig.~\ref{fig:tp}~(a) shows these websites with green, and it is interesting to note that these websites have higher than average number of cookies among those that provide cookie notices with choice.   
 
 Finally, Fig.\ref{fig:top500_cookietype} expands our study from the top 100 sites we have been looking at so far to the \texttt{Alexa.com} top 500 sites. As expected,  the fraction of sites offering users a choice drops drastically after the top 100. Many sites also close and stop serving EU users.

 \subsection{Cookie notices of top non-EU websites}

GDPR compliance is a requirement for all websites that wish to operate within or can be accessed from EU locations. Therefore, we are interested in understanding how non-EU websites have dealt with the introduction of GDPR as they will also be subject to the regulation if serving EU citizens in the EU. As mentioned previously, several prominent websites such as LATimes.com (\texttt{Alexa.com} rank 163 in the USA), Chicagotribune.com (\texttt{Alexa.com} USA rank 342) and  QQ.com (\texttt{Alexa.com} rank 2 in China), have once stopped serving users in the EU, serving up a banner that says they do not operate within EU boundaries because of GDPR. 

Therefore, as a baseline, we manually examine how \texttt{Alexa.com} top 100 sites in China and the USA serve cookie notices when accessed from the UK. Table~\ref{pie_cookie} shows the comparison of top 100 sites in the UK (also studied in Fig.~\ref{fig:tp}) and those in China (CN) and USA (US). In contrast with the UK, only 10\% (respectively 34\%) of sites in China (USA) offer users a choice of which cookies to store, and only a further 6\% (14\%) serve a cookie notice with no choice. Thus the vast majority (84\% in CN, 52\% in the US) of top sites are currently operating without a cookie notice. A large proportion also serve a notice that tracking cookies are being used, but users are not able to opt out of such cookies and continue to use the websites. Indeed, only a small fraction 10\% (34\%) of top sites  in CN (US) offer users a cookie notice with choice. Therefore, it appears that \emph{users of international non-EU websites in the UK obtain little protection, and little choice about their privacy and tracking}. 
\begin{table}[ht]
\centering
\small{\begin{tabular}{r|l|l|l}\hline
                      & UK   &  US  &   CN \\
Choice (UK)           & 42\%& 34\% &   10\%\\
Notice no Choice (UK) & 35\% & 14\% &   6\%\\
No Cookie Choice (UK) & 23\% & 52\% &  84\%\\\hline
\end{tabular}}
\caption{Percentage  distribution of different kinds of cookie notices in \texttt{Alexa.com} Top 100 websites from US, CN and UK.}\label{pie_cookie}
\vspace{-3mm}
\end{table}
\begin{figure}[ht]
\centering
\includegraphics[width=0.6\textwidth]{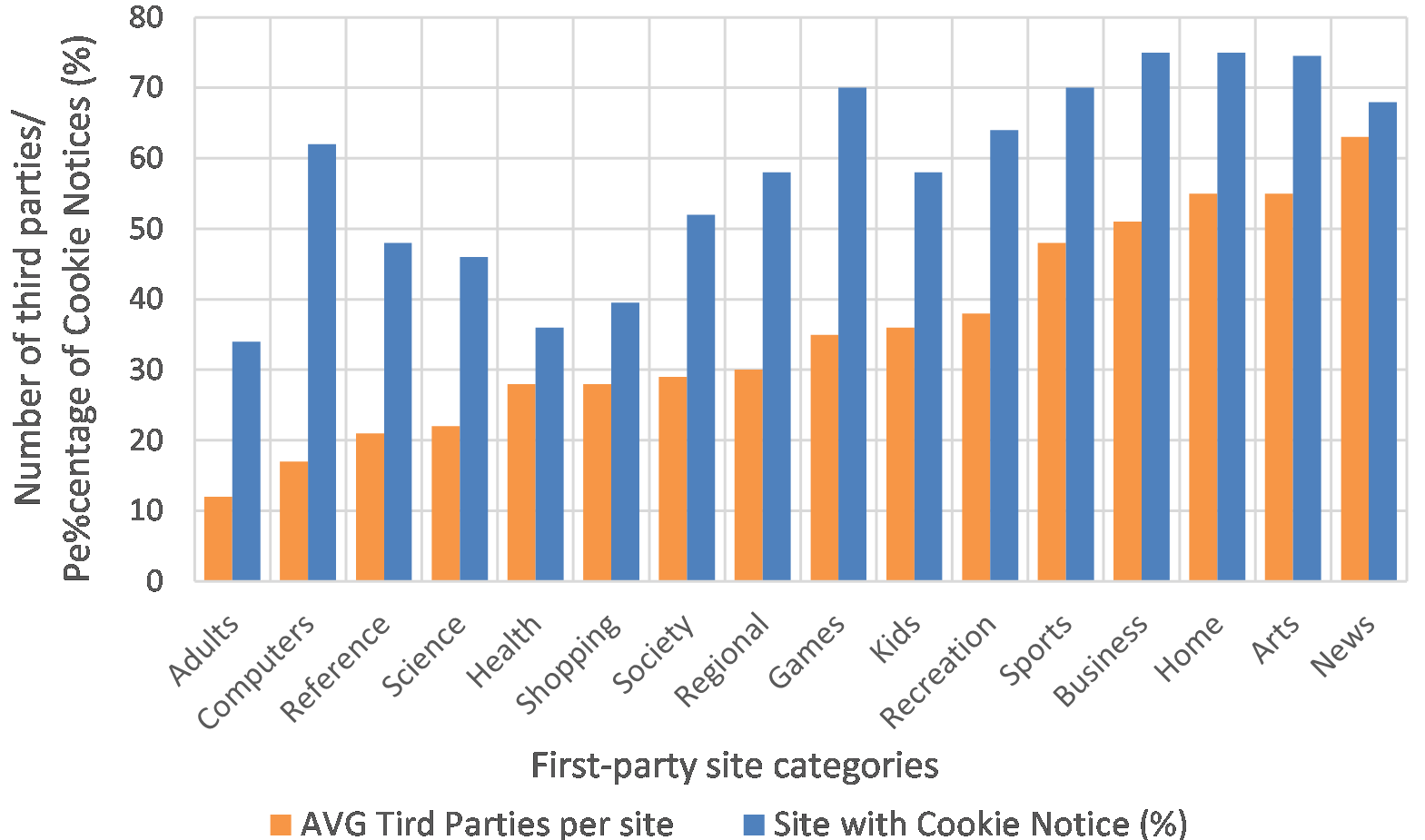}
\caption{\label{fig:Cate_cookie} The average number of third parties per site and percent of cookie notices in each category.}
\vspace{-3mm}
\end{figure}

We next turn to global top sites across categories in \texttt{Alexa.com}, to understand GDPR compliance among different kinds of websites. Fig.~\ref{fig:Cate_cookie} shows the categories ranked by the number of third parties per site for each category on average. The count in Adult websites is the least, likely because they typically are not able to access the most common third-party cookie providers such as Facebook or Google Analytics. However, Adult websites also have the lowest fraction of websites serving cookie notices. News and home related websites have the largest number of third parties, but also show the highest levels of compliance (i.e., serve cookie notices). In general however, no individual category of global websites achieves the same level of compliance as the top 100 UK websites.

\section{Cookie notices to real users}
\label{sec:users}
Until now, we have been studying how top sites around the web serve third-party cookie notices. However, any given user may have niche interests, and will likely access sites outside the list of \texttt{Alexa.com} top sites. To understand how compliant those less popular sites are, we turn to an ongoing user study we are conducting on third-party trackers collected by browser plugins, using a live user group. We also wish to understand whether real users see a decrease in number of tracking third party cookies after GDPR.

\noindent\textbf{Cookie notices in real users' browsing histories} We use 1528  websites collected by UK users in the weeks from Jan - Mar 2018 and evaluate the popularity of those sites by their visiting frequency to group them into 5 quintiles. 
Quintile 1 comprises 133 sites visited by over 80\% participants, quintile 2 has 150 sites  visited by around 60\% - 80\% users, quintile 3 is 148 sites visited by 40\% - 60\% users, quintile 4 168 sites by 20\% - 40\% users and quintile 5 has by far the most number of sites (929), but each site is visited by less than 20\% participants. Even the \texttt{Alexa.com} top 100 sites are evenly distributed across the five quintiles -- 15 of the \texttt{Alexa.com} UK top 100 sites fall in quintile 5, i.e., are visited by fewer than 20\% of users. 19 \texttt{Alexa.com} top100 sites are not accessed by \emph{any} user.



\begin{figure}[ht]
\centering
\vspace{-3mm}
\includegraphics[width=0.62\textwidth]{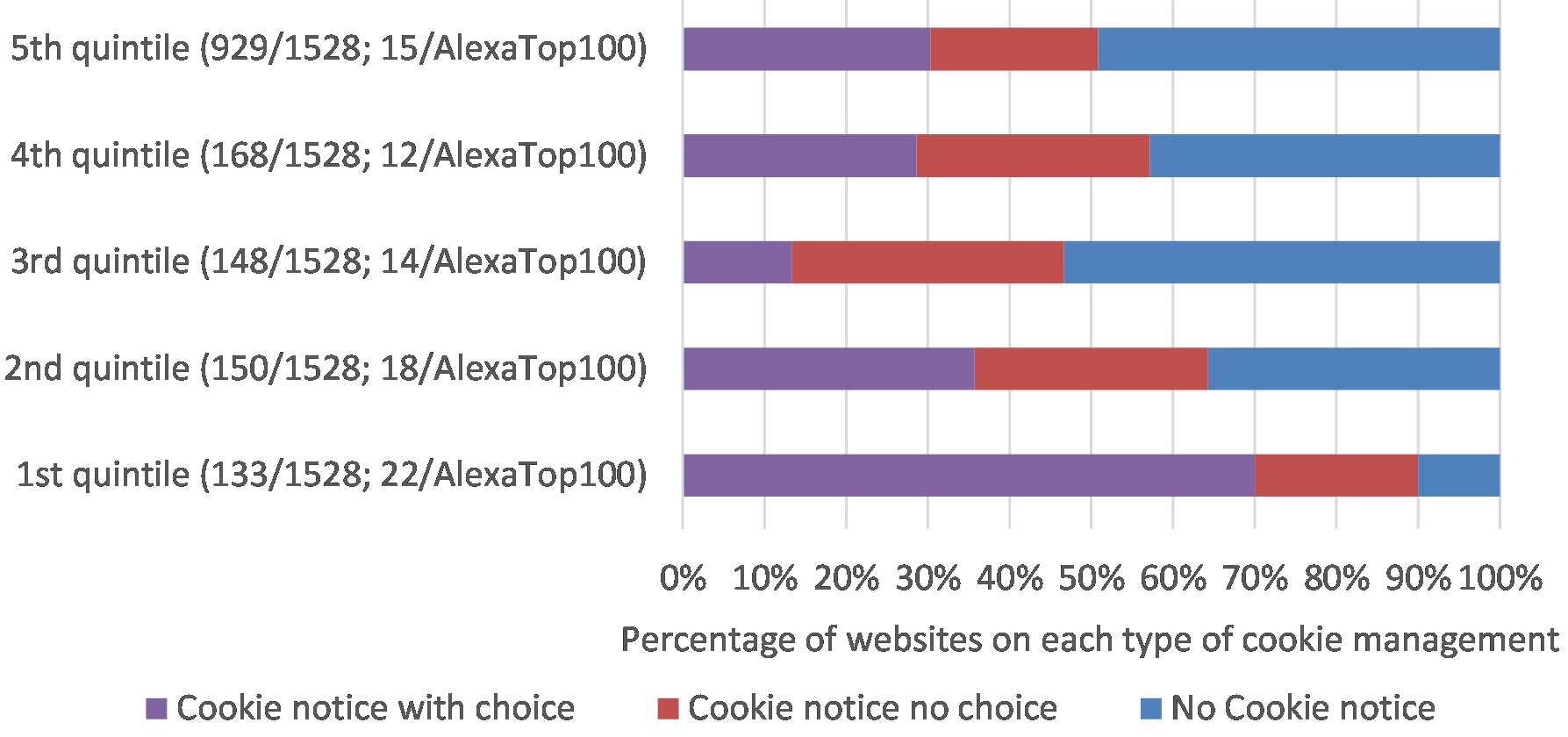}
\vspace{-2mm}
\caption{\label{fig:toprachel_cookie} Cookie notices among the five quintiles of websites accessed by a real user base.}
\vspace{-4mm}
\end{figure}

Fig.~\ref{fig:toprachel_cookie} shows the distribution of different kinds of cookie notices among the websites in different quintiles. Reassuringly, websites which are visited by most of the users in the study (quintile 1) has the highest fraction of websites which serve some form of cookie notice. However, as we go towards more niche interest websites, those visited by smaller numbers in our user study, the fraction that serve GDPR cookie notices drops drastically (there is a steady decline up to quintile 3, and although there is a brief uptick in quintiles 4 and 5, the fraction serving cookie notices are still below the top 2 quintiles). This suggests that users may need to be careful about niche websites.

\begin{figure}[!ht] 
\centering
\vspace{-2mm}
\includegraphics[width=0.6\textwidth]{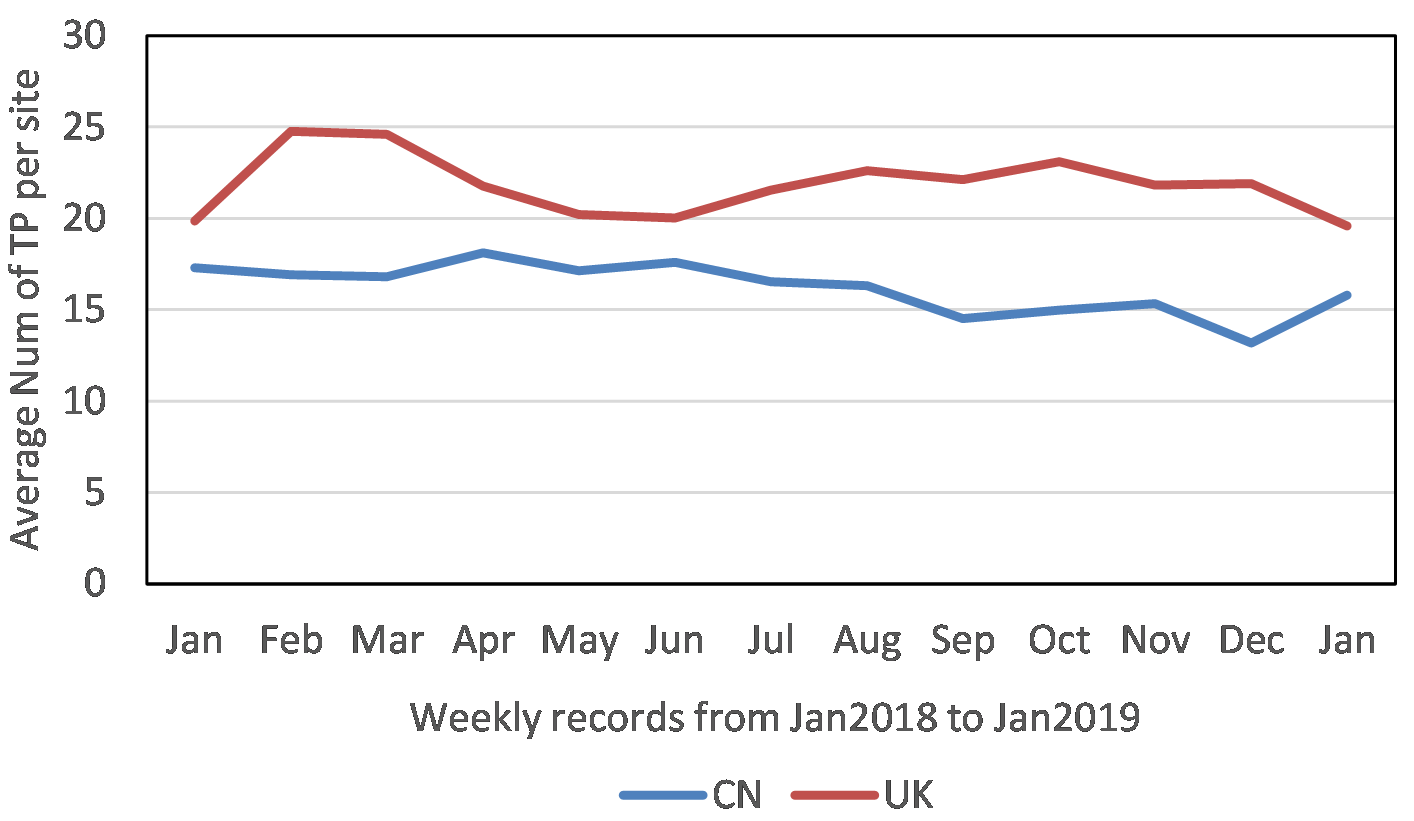}
\vspace{-2mm}
\caption{\label{fig:TP_yearly} Average number of third parties per site, based on weekly browsing records of UK and China participants.}
\vspace{-3mm}
\end{figure}

\noindent\textbf{Did GDPR affect third party cookie numbers for real users?} Whereas previous sections have looked at synthetic or programmatically generated browser visits to websites, we can also ask the \emph{extent to which users explicitly make use of the choice provided by GDPR cookie notices} and choose to block third-party tracking. We examine this using the anonymised cookie data from one year of browser histories of the UK users in our study. Fig.~\ref{fig:TP_yearly} shows that although there was a brief reduction in the number of third-party cookies when GDPR was introduced in May 2018, the overall number of cookies among the 9 UK users has stayed relatively the same between Jan 2018 and Jan 2019. The reductions between Mar 2018 and Jun 2018 appear to coincide with the beginning of the preparations for GDPR cookie compliance and the cookie consent manager rollouts of the widely used OneTrust \cite{onetrust_2018} (Mar 2018) and TrustArc \cite{TrustArc_2018} (Apr 2018)  for GDPR compliance, and similar reductions also reported by others\cite{libert_EUnews}. However, Do Not Track cookies and GDPR consent cookies expire; cookie caches get cleaned etc, and \emph{it appears that users in our study have subsequently mostly chosen default settings or have made choices that do not increase their privacy -- there is little change in the numbers of third-party cookies per website visited between early 2018 and early 2019}. Table~\ref{table:anonymous_sites7} shows how the numbers of cookies varied for selected sites of different \texttt{Alexa.com} ranks between Feb 2018 and Feb 2019, with a minimum being seen around the time GDPR introduced in May 2018. Interestingly, users in China experience \emph{fewer} third party cookies throughout the duration.

\begin{table}[ht]
\centering
\small{\begin{tabular}{c|lllll}
 & Site A (top100)  & Site B (top200) & Site C (top300) & Site D (top400) & Site E (top500)\\ \hline

Feb., 2018  & 13  & 14 & 20 & 21 & 37 \\
May, 2018   & 8 & 8 & 16 & 17 & 29 \\
Feb 2019   & 12  & 8 & 22 & 18 & 32 \\ \hline

\end{tabular}}
\caption{Number of cookies on websites visited by real users.}
\label{table:anonymous_sites7}
\vspace{-5mm}
\end{table}

\section{Related Work}
GDPR is newly introduced, and so there have only been a handful of measurements and analyses:  \cite{smaragdakis_tracking} concluded that tracking flows mostly stay within the EU. In a periodic survey of top 500 sites, \cite{degeling_ValuePrivacy}  found that around one-sixth of websites (15.7\%) had reorganised privacy policies by May 25, 2018. \cite{urban_CookieSync} investigated cookie synchronisation and show that GDPR cookie consents are insufficient to prevent leakages. Our work differs from these studies as we examine GDPR over a long duration, using real users' browsing histories and focusing on third-party cookies. 

Senzing Inc.~\cite{Missinglink_GDPR} suggests that around 60\% of European companies are not yet prepared for GDPR and 44\% of the EU's larges companies are worried about compliance with GDPR. \cite{vlajic_EUkids} studies kids and teenagers' privacy and finds the EU children may be subject to more third party tracking compared to the US. ~\cite{libert_EUnews} examines news websites and finds that UK in particular has a high level of tracking. Results such as these corroborate our findings that sites may not be offering a choice, or offering a choice and then not respecting users' choice (see examples from footnote 3).

Different from studying the behaviour of actual websites is to take an economic, policy or legal perspective. However, even in these fields, it is now being recognised that choice may be difficult for users to deal with, given the complexity of these sites and the technology used \cite{privacyrewiew_2016,privacy_economics}. \cite{libert_automated} develops a tool to examine privacy policies of websites to see if all third parties are being disclosed, and finds that privacy policies are extremely complicated, and several third parties are not being disclosed. Our results (Fig.~\ref{fig:TP_yearly}) also suggest that in practice users may not make choices that maximise privacy.

Our work focuses on GDPR consent cookies, but fits within the overall area of studying third-party tracking. Gomer~\textsl{et al.}~\cite{thirdpartytracking_2013} posited three key questions to explore: the relationship between search context and tracking services, the extent of tracking and the characteristics of tracking services.  We focus on the extent of tracking, and on the characteristics. \cite{trackingmeasure_2015} checked the coverage of top 10 trackers and showed that Doubleclick might cover over 80\%. Such works highlight the importance of third-party websites in leaking personal information and motivate our study of third-party cookies and consent notices about their use.





\section{Discussion and Conclusions}

In this paper, we took an in-depth look at the effect of GDPR, which requires cookie notices when sites are using third-party cookies that collect personal data. We find that although UK-based websites comply in general (i.e., serve some form of cookie notice), non-EU sites are less likely to offer fine-grained choices for users to decide their privacy preferences. Availability of choice also varies across different categories of websites, with adult websites being the least likely to offer a cookie notice, but also with many fewer third-party cookies than other categories such as news websites. 

Fine grained choices are not necessarily what is ``best'' for the users: First, though UK websites are meeting the cookie consent requirement by presenting users with a choice, this choice can be a false one -- if default choices are accepted, it could sometimes lead to higher numbers of third-party cookies than before. Second, by studying the numbers of third party cookies in real users' browsing histories, we find that GDPR has had little long term effect on the numbers of cookies. In practice, the choices, when offered, can be very fine grained (e.g., Fig.~\ref{cookie_linkedin}~(b)), allowing users to opt out of cookies from specific third parties that are being used by the website while still allowing them to opt in for cookies from other third parties. We speculate that users may be fatigued by the effort of having to choose their privacy preferences on every website they visit, and end up accepting the default choices offered by the websites (which in a majority of sites, is to have tracking turned on). Interestingly, users in the UK appear to have \emph{larger} numbers of third party cookies than countries like China. Unfortunately, tracking is the default on many sites where users are not given a choice at all, and the only real choice for users appears to be a forced one of either accepting  tracking and third party cookies, or   not using the website at all. 

In summary, we find that by and large, the relationship between website operators and users  remains unbalanced, and GDPR may in practice be falling short of the level of protection that it aims to deliver.

\bibliographystyle{unsrt}  



\end{document}